\documentclass[12pt]{article}

\usepackage{a4wide}
\usepackage{amssymb}
\usepackage{bm}
\usepackage{latexsym}
\usepackage{dcolumn}
\usepackage{amsfonts,amssymb}
\usepackage{graphicx,epsfig}
\usepackage{psfrag}
\usepackage{amsmath,amssymb}

\usepackage{bm}
\usepackage{latexsym}
\usepackage{mathrsfs}
\usepackage[T1]{fontenc}

\newcommand{\be}{\begin{equation}}
\newcommand{\ee}{\end{equation}}
\newcommand{\ba}{\begin{eqnarray}}
\newcommand{\ea}{\end{eqnarray}}
\newcommand{\ban}{\begin{eqnarray*}}
\newcommand{\ean}{\end{eqnarray*}}

\begin{document}
\begin{titlepage}
\pagestyle{empty}
\baselineskip=21pt
\vspace{2cm}
\begin{center}
{\bf {\Large 
How Emergent is Gravity? 
}}
\end{center}
\begin{center}
\vskip 0.2in
{\bf Swastik Bhattacharya} and {\bf S. Shankaranarayanan}
\vskip 0.1in
{\it School of Physics, Indian Institute of Science Education and Research 
Thiruvananthapuram (IISER-TVM), \\ 
Trivandrum 695016, India} \\
{\tt Email: swastik@iisertvm.ac.in, shanki@iisertvm.ac.in}\\
\end{center}

\vspace*{0.5cm}

\begin{abstract}
General theory of relativity (or Lovelock extensions) is a dynamical theory; given an 
initial configuration on a space-like hypersurface, it makes a definite prediction of the final 
configuration. Recent developments suggest 
that gravity may be described in terms of macroscopic parameters. It finds a concrete manifestation in the 
fluid-gravity correspondence. Most of the efforts till date has been to relate equilibrium configurations in 
gravity with fluid variables. In order for the emergent paradigm to be truly successful, it
has to provide a statistical mechanical derivation of how a given initial static configuration evolves into 
another. In this essay, we show that the energy transport equation governed by the 
fluctuations of the horizon-fluid is similar to Raychaudhuri equation and, hence gravity is truly emergent.
\end{abstract}

\vspace*{2.0cm}

\begin{center}
{\bf Essay for the Gravity Research Foundation essay competition in 2015} \\
{\bf Submitted on 30 March 2015}
\end{center}

\vfill\vfill

\end{titlepage}

\baselineskip=18pt

Emergent phenomena occur when simple interactions
working cooperatively create more complex interaction \cite{Anderson}.
Physically, simple interactions
occur at smaller length scales (microscopic level), and
collective behaviour manifests at much larger length scales. For example, 
the Coloumb force $(1/r^2)$ experienced by two charges
separated by a distance($r$) is understood to be a fundamental force while the
interaction force $(1/r^4)$ between two moving bubbles in a superfluid
is understood as an {\it emergent law}. In the same spirit, one is tempted to ask 
whether the General theory of relativity is the low-energy limit of a strongly correlated system and
gravitons their collective excitations \cite{Sakharov}.

Due to the long range and attractive nature of gravity, gravitational
systems are far-from-equilibrium. Over the last three
decades, it has been noticed that the condensed matter systems that are
far-from-equilibrium exhibit a convenient separation of length and
time scales \cite{Chaikin-Lubensky}.  
One of the key physical
ingredient is that the dynamics of a system with 
many degrees of freedom
can be described by the interaction of only a few (such as those at
long length and time scales).This so-called {\it hydrodynamic} approach 
provides a successful basis for describing systems far from equilibrium~\cite{Chaikin-Lubensky,Bettelheim}. 

The thermodynamics of black-holes \cite{BH-Thermo},
seem to suggest the {\it hydrodynamic} approach. Due to Hawking radiation, the
black-holes are out-of-equilibrium systems \cite{Hawking-1975,Landauer} 
and the black-hole entropy is
dominated by the degrees of freedom close to horizon~\cite{Shanki-Saurya-CQG-2005}. 
Over the last few years, a
growing body of evidence suggests that gravitational dynamics near the
black-hole horizon, is analogous to the dynamics of fluids~\cite{Shiraj,Paddy,Strominger}. (See also
Refs. \cite{Sakharov, Damour, Membrane,Jacobson,Verlinde}.)

Recently, the present authors have taken an alternative route and shown that the fluid-gravity 
correspondence is more physical and can be used to derive physical quantities  
from the horizon-fluid\cite{SSSB}. Identifying the long 
wavelength limit as that used in the Mean Field Theory description of Phase Transitions, we 
showed that the entropy of the ordered phase is same as the Bekenstein-Hawking entropy. The 
flow chart below provides a birds eye view of the fluid-gravity programme and the right side 
describes the programme undertaken by the authors. 

Mean Field Theory may yield correct critical exponents for certain phase transitions, 
however, it ignores fluctuations that drive system from one physical state to another. In the same 
spirit, while we were able to recover Bekenstein-Hawking entropy by modelling the horizon-fluid using Mean 
Field Theory, our previous analysis is incomplete, or for that matter, any emergent gravity approach
\cite{Paddy}. More specifically, the question that needs to be addressed in any emergent gravity approach 
is whether the fluctuations of the horizon-fluid provide necessary information about how a given 
horizon-fluid configuration evolves into another. In this essay, we show that the transport equation for 
the order parameter of the horizon-fluid system is identical to the Raychaudhuri equation. 

  \begin{center}
    \includegraphics[width=8cm]{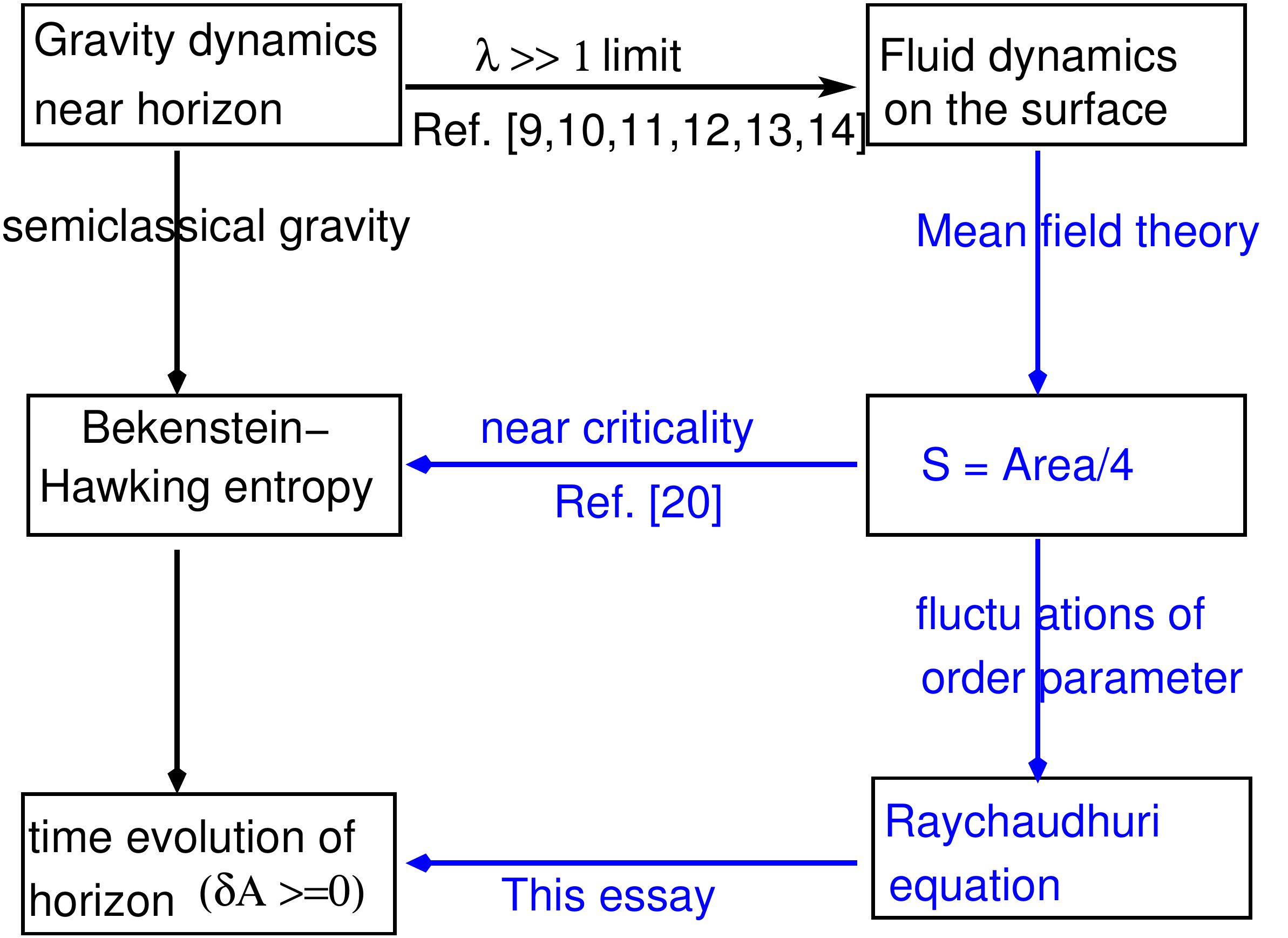}
   \end{center}

We start by assuming that the Horizon-fluid forms a condensate at 
a critical temperature. The justification comes from two lines of arguments. First, the 
evidence provided by Carlip\cite{Carlip} that the  black-hole horizon has some properties that exhibit 
universality. This indicates that the physics near the horizon is that of a system near a critical point.
Second, recently, Skakala and Shankaranarayanan\cite{Skakala} modelled the fluid as a Bose gas with 
$N$ particles and found that all the particles stayed in the ground state for large horizon radius. This 
suggests that horizon-fluid forms a BEC at some critical temperature
$T_c$. The conditions underlying this are \cite{SSSB}:
\begin{enumerate}
\item There is a temperature $T_c$ (critical temperature), at which,
  all the $N$ fluid degrees of freedom on the horizon form a condensate.
\item The system remains close to the critical point.
\end{enumerate}

One can describe this critical system, a homogeneous fluid, using Mean Field Theory.  
The order parameter is
\begin{equation}
 \eta= \sqrt{kN}. \label{eta}
\end{equation}
and the thermodynamic Potential $\Phi$ is %
\begin{equation}
 \Phi= \Phi_0+a(P)(T-T_c)\eta^2+B(P)\eta^4. \label{phiLT}
\end{equation}
where $k$ is a constant,  coefficients $a$ and $B$ are determined by the relation
$P=-TA$. One can determine the entropy of the fluid in  $\eta\ne0$
phase and show that it corresponds to the Bekenstein-Hawking entropy,
$S= \frac{A}{4}$ \cite{SSSB}. The Horizon-fluid is near a critical point and when
the system goes over to the ordered phase, its entropy is the same as
the black-hole entropy.  We showed that the formalism  
can be extended to include black-holes in AdS background\cite{SSSB}. The negative 
cosmological constant can be treated as an external field analogous to an 
external magnetic field. This
causes first order phase transition and the existence of a tri-critical point.

What has been described so far, concerns only static variables in the
gravity theory(in this case, the parameters which fully describe a
black-hole), correspond to the equation of
state of the horizon-fluid. Our aim now is to investigate, within the emergent paradigm, how a given static 
configuration with a given equation of state is driven to another static configuration. We do this in two 
steps. First, we consider fluctuations from the equilibrium that are adiabatic; fluctuations whose 
wavelengths are of the same order as the horizon-fluid and occur over a longer time scale. Second, 
we consider fluctuations that are non-adiabatic; wavelength of these fluctuations are usually smaller than 
the fluid scale.

{\it Adiabatic Process}:  Fluctuations take the fluid from one equilibrium configuration to another. 
Since these fluctuations are adiabatic, the new 
equilibrium configuration can also be described by Mean Field Theory. The equilibrium position is one of the 
two minimas of the double well potential on which the system has settled, i.e. 
$\eta_{min}^2= \frac{a(T_c-T)}{2B}$. Let, $\delta\eta_s$($\delta N_s$) denote the change in the value of the 
order parameter(number) due to fluctuations, then the change in the potential is
\begin{equation}
\delta\Phi_s= \frac{1}{4}\alpha (T-T_c)\frac{\delta N_s^2}{N_0}. \label{deltaphi}
\end{equation}
The change in entropy during this process can be
determined in two ways, 
$\delta S= -\frac{\partial \delta\Phi_s}{\partial T}$ and $\delta S= \frac{\partial S}{\partial N_s}$.\\
Comparing them, we get, $\delta E= T \delta S$. This is the statement of the First 
Law of Thermodynamics for the fluid system. It is of the same form as the mathematical statement of the 
process 1st law for event horizons,
\begin{equation}
\delta M= \frac{1}{4}T\delta A= T\delta S, \label{process1st}
\end{equation}
which relates the increase in the mass of a black-hole due to matter-energy falling through the horizon to the 
increase in the horizon area~\cite{Wald,Jacobson,Parentani}. It shows us that the physical 
process 1st law can be thought of as the adiabatic restoration of equilibrium after a fluctuation in the 
fluid system. 

{\it Non-Adiabatic Process:}
In this general case, the fluid system is away from equilibrium and some amount of
energy is being transferred to it from the external source. Due to the fluctuations, energy is 
gained by the horizon-fluid, correspondingly, the number of fluid d.o.f and order 
parameter $\eta$ change. 
Applying Onsager's hypothesis\cite{Onsager}, and assuming $k$ does not change in Eq.~\eqref{eta}, we can 
describe the change in the order parameter $\eta$ by the Langevin equation. 
We note that the Horizon fluid has negative Bulk viscosity. Since the fluid is taken to be homogeneous, only the bulk viscosity needs be 
considered. 

The change in the order parameter $\eta$ is given by the Brownian motion(See, for instance, Ref. 
\cite{Kittel}): 
\begin{equation}
 \ddot{\eta}= -\beta\dot{\eta}+ F(t), \label{BrE}
\end{equation}
where, $F(t)$ is the random term and $ \beta\dot{\eta}$ is the damping term due to the bulk viscosity of the 
fluid. Using \eqref{eta}, we  get,
\begin{equation}
 \frac{dx}{dt}= -\beta x-\frac{1}{2}x^2+ 2\frac{ F(t)}{\eta},  \label{Bei2}
\end{equation}
where, $x =\frac{\dot{N}}{N}$. $\eta\ne0$ here, as it fluctuates around $\eta= \eta_{min}\ne0$.
Taking the ensemble average on both sides of Eq. \eqref{Bei2} and using $\langle\frac{ F(t)}{\eta}\rangle=0$, we get,
%
\begin{equation}
 \frac{d\langle x\rangle}{dt}= -\beta\langle x\rangle-\frac{1}{2}\langle x\rangle^2-\frac{1}{2}\langle\Delta x\rangle^2, \label{Beav}
\end{equation}
where, $\langle\Delta x\rangle^2 (= \langle x^2\rangle-\langle x\rangle^2)$ is the mean squared 
fluctuation in $x$. \\
To determine $\langle x^2\rangle$, we note, 
$\overline{K.E.} =\frac{1}{2}\overline{\dot{\eta}^2}$ and $\overline{P. E.}= \frac{\alpha}{k}(T-T_c)\overline{\delta\eta_s^2}$, where, 
$\overline{X}$ denotes the time average of a quantity. Using Virial theorem\cite{Kittel}, we have, 
$\overline{K.E.} = \overline{P. E.}$. The fluctuations in $\eta$ are related to the dissipated energy density 
of the fluid $\rho_d$ implying that $\langle\dot{\eta}^2\rangle$ and $\langle x^2\rangle$ are related to 
$\rho_d$. Thus, we get,
\begin{equation}
  \frac{d\langle x\rangle}{dt}= -\beta\langle x\rangle-\frac{1}{2}\langle x\rangle^2+ 8k\alpha\rho_d. \label{BrRC}
\end{equation}
The damping coefficient $\beta$ has dimensions of length-inverse. For the horizon-fluid, this corresponds to the 
horizon radius, which is inversely proportional to $T$. This leads to:
\begin{equation}
  \frac{d\langle x\rangle}{dt}= C \, T \, \langle x\rangle \, - \, \frac{1}{2}\langle x\rangle^2 \, + \, 8k\alpha\rho_d \, 
  \label{BrRC1x} 
\end{equation} 
where, $C$ is a constant. 
This is the key result of this essay regarding which we would like to stress 
the following points: 
\begin{enumerate}
 \item The above equation is similar to the Raychaudhuri equation sans the shear term, 
\begin{equation}
 \frac{d\theta}{dt}= 2\pi T\theta-\frac{1}{2}\theta^2-8\pi T_{\alpha\beta}\xi^\alpha\xi^\beta. \label{RC}
\end{equation}
To relate \eqref{BrRC1x} to the gravity side, we shift to the geometric variable, $A$, which is the cross-sectional area 
of the null congruence in addition to being the area of the $2+1$ dimensional fluid. Then 
$x=\frac{\dot{A}}{A}= \theta$. Here $t$ denotes the affine parameter along the null congruence as used in 
\cite{Damour}. Comparing \eqref{BrRC1x} with \eqref{RC}, one can write, $C=-2\pi$
and $\rho_d=\frac{1}{8k\alpha}(8\pi T_{\alpha\beta}\xi^\alpha\xi^\beta+\langle\dot{\eta}\rangle^2)$. 
It is to be noted that except for the values of the
coefficients, the signatures are the same. The Bulk
viscosity is negative for the fluid on the horizon. Hence, what is
normally the damping term in a standard Brownian motion, here reinforces the transport of energy. It
is reasonable to assume that the coefficient of such a term is
proportional to the Bulk Viscosity, $\eta_B$ (Stokes' law is one
example of this), i.e. $\beta= -CT$.  

\item The dissipated energy density is of the
form, $T_{\alpha\beta}\xi^\alpha\xi^\beta+
\frac{1}{2}\langle\dot{\eta}\rangle^2$. So there is an extra term
apart from the amount of the matter-energy that falls across the
horizon. In the gravity picture, the dissipated energy then consists
of two parts, one matter-part and one geometry-part. Thus, our analysis provides 
an alternative view of the Raychaudhuri equation i. e.  governing the
transport of energy on the fluid side. 
\end{enumerate}

General Theory of Relativity is a 
dynamical theory and makes a clear and precise prediction about how an initial configuration would evolve to a 
final configuration. Raychaudhuri equation is a crucial ingredient in describing the dynamics of gravity. It 
governs the dynamics of null geodesic congruences. The emergent gravity paradigm has been with us for close  
to fifty years. Most of the efforts within the emergent gravity approach has been directed towards 
relating the equilibrium configurations in gravity with the fluid variables. In this essay, we have shown 
that the differential equation governing the transport of energy into the fluid is similar to the 
Raychaudhuri equation. This clearly demonstrates that the Horizon-Fluid description provides dynamical 
information about gravity and gravity is truly emergent. One might compare this with the long wavelength or 
the hydrodynamic limit of AdS-CFT, where the dynamics of the fluctuations on the fluid side can be mapped to 
the macroscopic variables on the gravity side\cite{Son,Shiraj}. Such a programme, however, has not been 
carried out for the gravity theory for a more general class of space-time. The results reported here 
constitute a first step towards that direction.

\section*{Acknowledgments}

The work is supported by Max Planck-India Partner Group on Gravity and
Cosmology. SS is partially supported by Ramanujan Fellowship of DST,
India.

\end{document}